# A diverse set of two-qubit gates for spin qubits in semiconductor quantum dots

Ming Ni,[1,2] Rong-Long Ma,[1,2] Zhen-Zhen Kong,[3] Ning Chu,[1,2] Sheng-Kai Zhu,[1,2] Chu Wang,[1,2] Ao-Ran Li,[1,2] Wei-Zhu Liao,[1,2] Gang Cao,[1,2,6] Gui-Lei Wang,[3,5,6] Guang-Can Guo,[1,2] Xuedong Hu,[4] Hai-Ou Li,[1,2,6,*] and Guo-Ping Guo[1,2,6,7,*]

[1] *CAS Key Laboratory of Quantum Information, University of Science and Technology of China, Hefei, Anhui 230026, China*

[2] *CAS Center for Excellence in Quantum Information and Quantum Physics, University of Science and Technology of China, Hefei, Anhui 230026, China*

[3] *Integrated Circuit Advanced Process R&D Center, Institute of Microelectronics, Chinese Academy of Sciences, Beijing 100029, P. R. China*

[4] *Department of Physics, University at Buffalo, SUNY, Buffalo, New York 14260, USA*

[5] *Beijing Superstring Academy of Memory Technology, Beijing 100176, China*

[6] *Hefei National Laboratory, University of Science and Technology of China, Hefei 230088, China*

[7] *Origin Quantum Computing Company Limited, Hefei, Anhui 230026, China*

*Corresponding authors: haiouli@ustc.edu.cn; gpguo@ustc.edu.cn.

**To realize large-scale quantum information processes, an ideal scheme for two-qubit operations should enable diverse operations with given hardware and physical interaction. However, for spin qubits in semiconductor quantum dots, the common two-qubit operations, including CPhase gates, SWAP gates, and CROT gates, are realized with distinct parameter regions and control waveforms, posing challenges for their simultaneous implementation. Here, taking advantage of the inherent Heisenberg interaction between spin qubits, we propose and verify a fast composite two-qubit gate scheme to extend the available two-qubit gate types as well as reduce the requirements for device properties. Apart from the formerly proposed CPhase (controlled-phase) gates and SWAP gates, theoretical results indicate that the iSWAP-family gate and Fermionic simulation (fSim) gate set are additionally available for spin qubits. Meanwhile, our gate scheme limits the parameter requirements of all essential two-qubit gates to a common $J{\sim}\Delta E_Z$ region, facilitate the simultaneous realization of them. Furthermore, we present the preliminary experimental demonstration of the composite gate scheme, observing excellent match between the measured and simulated results. With this versatile composite gate scheme, broad-spectrum two-qubit operations allow us to efficiently utilize the hardware and the**

**underlying physics resources, helping accelerate and broaden the scope of the upcoming noise intermediate-scale quantum (NISQ) computing.**

**Main text.**

The study of quantum computing has experienced tremendous progress over the past two decades, such that a variety of qubit platforms, from superconducting qubits and trapped ion qubits, to spin qubits in silicon more recently, are now in the Noisy Intermediate-Scale Quantum (NISQ) regime [1-10]. As quantum circuits grow in size, developing optimal quantum gate sets is emerging as a major challenge toward the building of a universal quantum computer.

A minimal gate set for universal quantum computing consists of arbitrary single-qubit rotations and a single two-qubit entangling gate [11]. Availability of a larger variety of two-qubit gates could help improve the quantum circuit depth and ensure that the hardware resources are utilized more efficiently [12, 13]. Furthermore, specially designed quantum simulations [14], algorithms [15] and error mitigation techniques [16] may be implemented more efficiently using specific gates such as the Fermionic simulation (fSim) gate set [14, 17]. On the other hand, while a given physical interaction may be used to generate any two-qubit gate, each two-qubit gate has a native interaction that can generate it efficiently, essentially by just turning on the interaction for a certain period of time. For example, a controlled phase shift (CPHASE gate) can be generated via Ising interaction; a controlled rotation (CROT) via Ising interaction and a selected ESR/EDSR pulse; a SWAP gate (or $\sqrt{\text{SWAP}}$) via Heisenberg exchange; and an iSWAP gate via XY interaction. In other words, given a particular physical interaction in a qubit system, usually one specific two-qubit entangling gate can be generated efficiently, while any other two-qubit gate would have to be obtained via a more complicated pulse sequence, making these gates available but less efficient [18-21].

Spin qubits in semiconductor quantum dots (QDs, especially in Si and Ge), are one of the promising platforms for quantum computing, and are now in the early NISQ era [7-10, 22-24]. The native interaction for spin qubits is the Heisenberg exchange interaction, $J\vec{S}_1 \cdot \vec{S}_2$, which generates SWAP (exchanging the states of two qubits) and $\sqrt{\text{SWAP}}$ gates efficiently if the two qubits are in resonance [25-27], with the former allowing on-chip qubit transportation [28], while a pair of the latter together with single-qubit gates leading to a controlled-NOT (CNOT) gate [18]. However, since single-spin operations are usually based on a magnetic field gradient produced by an on-chip micromagnet [29, 30], two neighboring spins are often off resonance by design, making SWAP gates much more difficult to generate, requiring reliable and precise diabatic tuning of exchange coupling $J$ and Zeeman energy difference $\Delta E_Z$ between the two qubits [26, 27]. Instead, CPHASE and CROT gates [31-33] are the main entangling gates for current generation spin qubits. Furthermore, iSWAP gate family has not been demonstrated for spin qubits at all. How to diversify the two-qubit gate set under the given constraints (exchange interaction together with magnetic field gradient) remains an intriguing open question for spin qubits.

We propose and verify a diabatic composite two-qubit scheme to implement a diverse set of two-qubit gates, including five types of essential two-qubit gates, iSWAP-family gates, and the fSim gate set, with reduced device properties requirements. We extend the basic diabatic two-qubit operations theory to the composition scenario, and propose our diabatic composite gate scheme. The simulation results show that the composite gate can separately tune the CPhase operation angle and iSWAP operation angle, thus facilitating the realization of diverse two-qubit operations. We calculate the maximum $J/\Delta E_Z$ required for five types of essential two-qubit gates, observe that the parameter requirements of them fall within a common $J \sim \Delta E_Z$ region. In experiments, we present the preliminary experimental demonstration of the composite gate scheme in a purified Si-MOS double quantum dots (DQD) device. Comparing the measured results to the theoretically predicted ones, excellent match between them verifies that our composite two-qubit gate scheme can simultaneously enabled a diverse set of two-qubit gate operations in an easily available parameter region.

**Results.**

**Diabatic Two-qubit Operation.** A schematic diagram of our two-qubit device is shown in Fig. 1(a), which is identical to that reported in previous works [26, 34, 35]. A two-qubit gate can be performed by adjusting the tunnel coupling $t_c$ and/or detuning $\varepsilon$ [36]. When a diabatic pulse is applied to modify $\varepsilon$, turning on the exchange interaction between the two qubits, the two-qubit gate operator in the two-spin basis of $(|\uparrow\uparrow\rangle, |\uparrow\downarrow\rangle, |\downarrow\uparrow\rangle, |\downarrow\downarrow\rangle)^T$ can be written as (see Supplementary Note 1 for detail)

$$U = Z_L(\gamma_L)Z_R(\gamma_R)\begin{pmatrix} 1 & 0 & 0 & 0 \\ 0 & \cos\left(\frac{\phi_0}{2}\right) - i\cos(\theta_0)\sin\left(\frac{\phi_0}{2}\right) & -i\sin(\theta_0)\sin\left(\frac{\phi_0}{2}\right) & 0 \\ 0 & -i\sin(\theta_0)\sin\left(\frac{\phi_0}{2}\right) & \cos(\phi_0) + i\cos(\theta_0)\sin\left(\frac{\phi_0}{2}\right) & 0 \\ 0 & 0 & 0 & e^{-i\eta_0} \end{pmatrix}, \quad (1)$$

where $\gamma_L$ and $\gamma_R$ are single-qubit local phases, which we do not focus on in this work. The Heisenberg exchange interaction contains both an Ising interaction term and an XY interaction term. Combined with the Zeeman energy difference between qubits, the diabatic pulse executes a two-qubit gate that can be decomposed into two parts: One is a CPhase operation due to the Ising interaction, with a rotating angle $\eta_0 = Jt$, where $t$ is the duration of the diabatic pulse [33]. The other part, as shown in Fig. 1(b), is a rotation operation in the subspace of the antiparallel states, produced by the combined effect of XY interaction and the non-zero $\Delta E_Z$. This operation introduces a rotation with an accumulated phase of $\phi_0 = \Delta Et$ around a tilted axis with $\theta_0 = \arctan\left(\frac{J}{\Delta E_z}\right)$ [37], where $\Delta E$ is the energy difference between the antiparallel states $|\widetilde{\uparrow\downarrow}\rangle$ and $|\widetilde{\downarrow\uparrow}\rangle$.

Figure 1(c) shows $J$, $\Delta E$ and $\Delta E_z$ at different $\varepsilon$, and Fig. 1(d) gives the corresponding three operation angles $\phi_0$, $\theta_0$, and $\eta_0$. With increasing $\varepsilon$, $J$ increases while $\Delta E_z$ decreases due to the Stark shift [38, 39]. Here we define $\varepsilon = 0$ at the symmetric operation point (point C in Fig. 1(c) insert) [36] and positive $\varepsilon$ corresponds to higher energy for $Q_R$. We calculate $\eta_0$ and $\theta_0$ while guaranteeing the operation duration $\tau_s$ making $\phi_0 = \pi$ at different $\varepsilon$. With $\varepsilon$ increases, $\theta_0$ and $\eta_0$ increase as well, while $\eta_0$ will gradually converge to $\phi_0$, due to the approached $J$ and $\Delta E$. At the detuning where $J(\varepsilon) = \frac{\sqrt{3}\Delta E_z(\varepsilon)}{3}$, a diabatic CZ gate can be realized since $\eta_0 = Jt = \pi$ and $\phi_0 = 2\pi$ will occur simultaneously, during which only the Ising interaction part remained. At the regime $J \gg \Delta E_z$, a CPhase operation with $\eta_0 = \pi$, and an iSWAP operation with $\phi_0 = \pi, \theta_0 = \frac{\pi}{2}$ are realized simultaneously here. The two parts of two-qubit operations introduced by the native exchange interaction gives the combined SWAP gate together.

Although a single-segment diabatic pulse can implement a two-qubit CPhase gate or a SWAP gate,

there is always a strict requirement on the ratio of $\frac{J}{\Delta E_z}$. When this parameter requirement cannot be satisfied, the iSWAP operation and the CPhase operation are always implemented simultaneously and inevitably mixed, thus the combined two-qubit operation will not be any of the essential two-qubit gates. And this final gate operation is the results of mixed effect of the Zeeman energy difference, plus Ising interaction and XY interaction in the native exchange interaction. Although we know the speed of these two operations accurately, we are still limited by the need to adjust the operation angles, since we cannot control their relative rates effectively and separately. If the operation angles of the iSWAP operation and the CPhase operation can be controlled separately, a diverse set of combined two-qubit gates will be available, and the exchange interaction between the spins can be sufficiently utilized.

**Composite Gate Scheme.** To overcome the difficulties associated with a single-segment diabatic pulse, we introduce a versatile gate implemented by a three-segment composite pulse, which can separately control the operation angle of the CPhase operation and iSWAP operation. As shown in Fig. 2(a), for a composite pulse composed of three continuous diabatic square pulses, we adjust $\eta_i$ and $\theta_i$ by changing the amplitude of each segment as well as having each segment implementing an operation with $\phi_i = \pi$ by changing the pulse durations. The total operator of this composite pulse (see Supplementary Note 1 for a detailed theoretical derivation) takes the form

$$U = U_3 U_2 U_1$$
$$= Z_L\left(\gamma_L + \frac{\pi}{4}\right) Z_R\left(\gamma_R - \frac{\pi}{4}\right)$$
$$\begin{pmatrix} 1 & 0 & 0 & 0 \\ 0 & \cos(\theta_1 - \theta_2 + \theta_3) & -i\sin(\theta_1 - \theta_2 + \theta_3) & 0 \\ 0 & -i\sin(\theta_1 - \theta_2 + \theta_3) & \cos(\theta_1 - \theta_2 + \theta_3) & 0 \\ 0 & 0 & 0 & e^{-i\pi[\sin(\theta_1) + \sin(\theta_2) + \sin(\theta_3)]} \end{pmatrix} \quad (2)$$
$$Z_L\left(-\frac{\pi}{4}\right) Z_R\left(\frac{\pi}{4}\right),$$

where the indices denote the three segments, and $\gamma_L$ and $\gamma_R$ are the single-qubit local phases of the two qubits. Analogous to the single-segment diabatic pulse, the three-segment pulse also consists of three components, including a single-qubit rotation, a two-qubit CPhase operation, and a rotation in the two-qubit subspace always with $\theta = \frac{\pi}{2}$. In this paper, we will focus on the two-qubit gate operations. Figure 2(b) shows the evolution trajectory of the qubit states while applying a composite pulse. Specifically, the CPhase operation angle $\eta = [\sin(\theta_1) + \sin(\theta_2) + \sin(\theta_3)]\pi$ is the superposition of the CPhase operations of each segment, as the consequence of the Ising component of the exchange interaction. In addition, the superposition of the rotation operations in the two-qubit subspace leads to a two-qubit iSWAP operation with a rotation angle $\phi = 2(\theta_1 + \theta_2 + \theta_3)$. This operation is the combined effect of Zeeman energy difference and XY interaction. Benefiting from $\phi_i = \pi$, the tilt axis angle is always equal to $\frac{\pi}{2}$, hence an exact iSWAP operation can always be realized. Due to the composite pulse scheme, now the two two-qubit operation angles can be modified by changing $(\theta_1, \theta_2, \theta_3)$ separately. Such a composite scheme is not limited to three segments. For the scenario in which each pulse segment realizes a $\pi$-rotation, the demonstrated composite two-qubit operation can always be given as

$$\eta = \pi \sum_{n=1,2,3\cdots}^{n} \sin(\theta_n), \tag{3}$$

$$\phi = 2\left[\sum_{n=1,3,5\cdots}^{2m-1} \theta_n - \sum_{m=2,4,6\cdots}^{2m} \theta_n\right]. \tag{4}$$

where, n is the number of pulse segments.

Without loss of generality, here we focus on the three-segment scenario for simplicity. By adjusting the amplitude of each segment of the composite pulse, we can control $J$ and $\Delta E_Z$ in order to modify $\theta_1$, $\theta_2$, $\theta_3$, which would in turn allow us to control the two two-qubit operations via their operation angles. Further control can be achieved via tunnel coupling $t_c$. Simulation results are given in Fig. 2(c). Assuming $\theta_1$, $\theta_2$ and $\theta_3$ can be adjusted from 0 to $\frac{\pi}{2}$ (the maximum tunable region), we give the surfaces corresponding to $\eta = n\pi$ and $\phi = \frac{n\pi}{2}$. In Fig. 2(c), we also highlight the intersections among these surfaces and indicate the essential two-qubit gate operations represented by each crossing curve. Interestingly, multiple types of essential two-qubit operations can be realized. For example, a $\sqrt{SWAP}$ is realized when $\eta = \frac{3\pi}{2}$ and $\phi = \frac{\pi}{2}$, while a $\sqrt{iSWAP}$ is realized when $\eta = 2\pi$ and $\phi = \frac{\pi}{2}$.

With knowledge of the dependencies between the operations $(\eta, \phi)$ and the rotation angles $(\theta_1, \theta_2, \theta_3)$, we deduce the available operation angle scale and the corresponding parameter requirements. Figure 3 gives the available CPhase operation angle as a function of the iSWAP operation angle. In the region $\phi \in \left[0, \frac{\pi}{2}\right]$, we see that $\eta$ can be tuned continuously for more than $2\pi$, which indicates that the operation angles of these two types of two-qubit operations can be controlled continuously and separately. The corresponding lower bound of the required rotation angle $\theta_{lb}$ is given in Fig. 3 as well as the corresponding requirements on $\frac{J}{\Delta E_Z}$ for five types of essential two-qubit gates. Most of the essential two-qubit operations can be demonstrated directly in the region where $J \sim \Delta E_Z$, excluding the SWAP operation, which can be realized by combining a CPhase gate and an iSWAP gate. In addition to the essential two-qubit gates, while maintaining $\eta = 2\pi$, the iSWAP-family gates can be realized by changing $\theta$. The fSim gate set [14, 17] is also feasible because of the continuously tunable operation angle. The high degree of tunability and versatility of the composite pulse allows us to impose some constraints for convenience. Considering the symmetric effect of the first and third pulse segments, in the analysis below we make a reasonable simplification, maintaining $\theta_1 = \theta_3$, as well as the amplitudes of the first and third segments are the same.

**Experimental Verification.** To experimentally verify the composite gate scheme, we measure the final state after implementing the composite operations on a two-qubit system. Specifically, the left qubit $Q_L$ is read out and initialized directly with Elzerman readout (at point A in Fig. 1(c) insert), while the right qubit $Q_R$ is read out indirectly by first shuttling the electron to the left quantum dot, then performing Elzerman readout (at point B in Fig. 1(c) insert). Under an external magnetic field of $B_0 = 605$ mT, we characterize $J(\varepsilon)$, $\Delta E(\varepsilon)$ and $\Delta E_z(\varepsilon)$ by measuring the DQD spectrum, as well as the dependence $\theta(\varepsilon)$ for a single-segment diabatic pulse. With a given $\theta(\varepsilon)$, we can then obtain the composition operation angle $(\eta, \phi)$ with the varied three-segment pulse amplitude $(\varepsilon_1, \varepsilon_2, \varepsilon_3)$, as given in Fig. 4(a-b). The theoretically predicted $P_{L,|\uparrow\rangle}$ is subsequently shown in Fig. 4(c) with the known operation angle (details in Supplementary Note 1).

In our experiments, after initializing the two qubits to the $|\uparrow\downarrow\rangle$ state, we apply a composite pulse with amplitudes $(\varepsilon_1, \varepsilon_2, \varepsilon_3)$, then read out the spin-up probability of $Q_L$ $P_{L,|\uparrow\rangle}$. The measured $P_{L,|\uparrow\rangle}$ as a function of $\varepsilon_1(\varepsilon_3)$ and $\varepsilon_2$ is shown in Fig. 4(d), which agrees excellently with the simulated result. This agreement verifies the validity of our composite operation scheme. For the pulse parameters on the contour line $P_{L,|\uparrow\rangle} = 0$, the implemented composite pulse includes an iSWAP gate and a CPhase operation with a nonzero rotation angle $\eta$. For the pulse parameters on the contour lines $P_{L,|\uparrow\rangle} = 0.5$, the implemented composite pulse includes a $\sqrt{iSWAP}$ gate and a CPhase operation.

In our device, due to the difference between the g-factors, $\Delta E_z$ increases with increasing $B_0$. To characterize the composite gate when $J \sim \Delta E_z$, we increase $B_0$ to 755 mT, where the variation interval of $\Delta E_z(\varepsilon)$ is 23~15 MHz and the variation interval of $J(\varepsilon)$ is 0~60 MHz. After calculate the composition operation angle $(\eta, \phi)$ with the varied three-segment pulse amplitude $(\varepsilon_1, \varepsilon_2, \varepsilon_3)$ (see Supplementary Note 2), we can simulate the measurement $P_{L,|\uparrow\rangle}$ at $B_0 = 755$ mT, as given in Fig. 4(e). The corresponding experimental results are shown in Fig. 4(f). The great agreement between the simulated and experimental results indicates that the composite pulses can reliably perform desired operations, including flipping spin states far away from the $J \gg \Delta E_z$ region.

**Discussion.**

Based on this composite gate framework, further dynamically corrected gates with better noise resistance can potentially be developed [40-46]. We believe that additional degrees of freedom, such as the pulse duration of each segment and the number of pulse segments, still need to be theoretically and experimentally researched. We expect that there exists a composite operation with a better extended parameter limitation compared to the basic situation in our scheme. In addition, more precise and comprehensive experimental research on composite gates should be conducted. Although, in Supplementary Note 3, the theoretically calculated results indicate the potential of high-fidelity composite gates, the quality of composite gate operation still needs to be evaluated experimentally, and benchmarking methods such as unitary tomography [17] and XEB [47] can be favorable solutions. To implement the composite gate in a large qubit array, we believe that feedback and device stability improvement are crucial, and a high circuit bandwidth and precise predistorted pulse are also necessary. We suppose that the state-of-the-art techniques [48-52] have been suitable for these requirements. With this composite operation scheme, we believe that the NISQ era will be accelerated, benefiting from its continuous operation angle and extended parameter limitations.

In summary, we propose and verify an extremely versatile composite two-qubit gate scheme for spin qubits coupled via the Heisenberg exchange interaction. By controlling the amplitude of the three segments, we can realize all the important two-qubit gates, including SWAP, iSWAP, CPhase, and fSim gates, even in the presence of a finite Zeeman energy difference $\Delta E_Z$ between the two qubits. We have experimentally verified the scheme in the region $J \sim \Delta E_Z$ on a purified Si-MOS DQD device. With greatly enhanced two-qubit operation flexibility, reduced parameter limitations, and increased operation speed, the diverse set of two-qubit gates enabled by our composite two-qubit scheme is potentially extensively utilized for realizing quantum algorithms with limited hardware and physical resources. In the future, device structure design and quantum circuit compilations can be simplified, and a wide variety of quantum algorithms and quantum simulations based on these composite gates can be applied to silicon quantum dots.

## Methods

**Device structure and fabrication.**

Our device is fabricated based on a 70 nm $^{28}$Si epilayer with a residual $^{29}$Si concentration of 60 ppm. As shown in Fig. 1(a), the two qubits $Q_L$ and $Q_R$ are located underneath the two plunger gates PL and PR, respectively. The charge occupancy is measured by the SET conductance, which capacitive couples with the DQD. A Ti/Co/Ti rectangular magnet is fabricated beside the quantum dots to provide the gradient magnetic field to drive the single-qubit oscillation with EDSR.

**Experimental setup.**

We performed the experiments in an Oxford Triton dilution refrigerator at a base temperature of 20 mK. After I/Q modulation, two microwaves from the output of the vector microwave source E8267D are generated to drive the single-qubit rotation. Two pulses channels of arbitrary waveform generator Tektronix 5208 are utilized to exert the diabatic composite pulses. The microwaves and the diabatic pulses are combined together at room temperature and exerted on the plunger gates through high-bandwidth control line. The initialization and readout pulses are generated by arbitrary waveform generator Tektronix 5204, combined with the DC voltage at room temperature, and exerted on the plunger gates through the low-frequency control line. The read out SET current, after converted to voltage through the IV converter DLPCA-200, is then filtered and amplified at room temperature and sampled using an Alazar digital card.

## Data availability

All the data that support the findings of this study are available from the corresponding author upon reasonable request.

## Acknowledgements

We acknowledge Chengxian Zhang for helpful discussions of theoretical derivation. This work was supported by the National Natural Science Foundation of China (Grants No. 12074368, 92165207, 12034018 and 92265113), the Innovation Program for Quantum Science and Technology (Grant No. 2021ZD0302300), the Anhui Province Natural Science Foundation (Grants No. 2108085J03), and the USTC Tang Scholarship. X. H. acknowledge financial support by U.S. ARO through Grant No. W911NF2310018. This work was partially carried out at the USTC Center for Micro and Nanoscale Research and Fabrication.

## Author contributions

M. N. develop the theoretical derivation with the help of X. H, and S.-K. Z. M. N. and H.-O. L perform the bulk of measurement and data analysis. R.-L. M. fabricated the device. Z.-Z. K. and G.-L. W. supplied the purified silicon substrate. M. N. wrote the manuscript with inputs from other authors. G.-C.G. advised on experiments. C. W., A.-R. L., N. C. and W.-Z. L. contributed to the simulation and G. C. polished the manuscript. H.-O. L. and G.-P. G. supervised the project. All the authors contributed to discussions.

## Competing interests

All authors declare that they have no competing interests.

# References.

# Figure captions

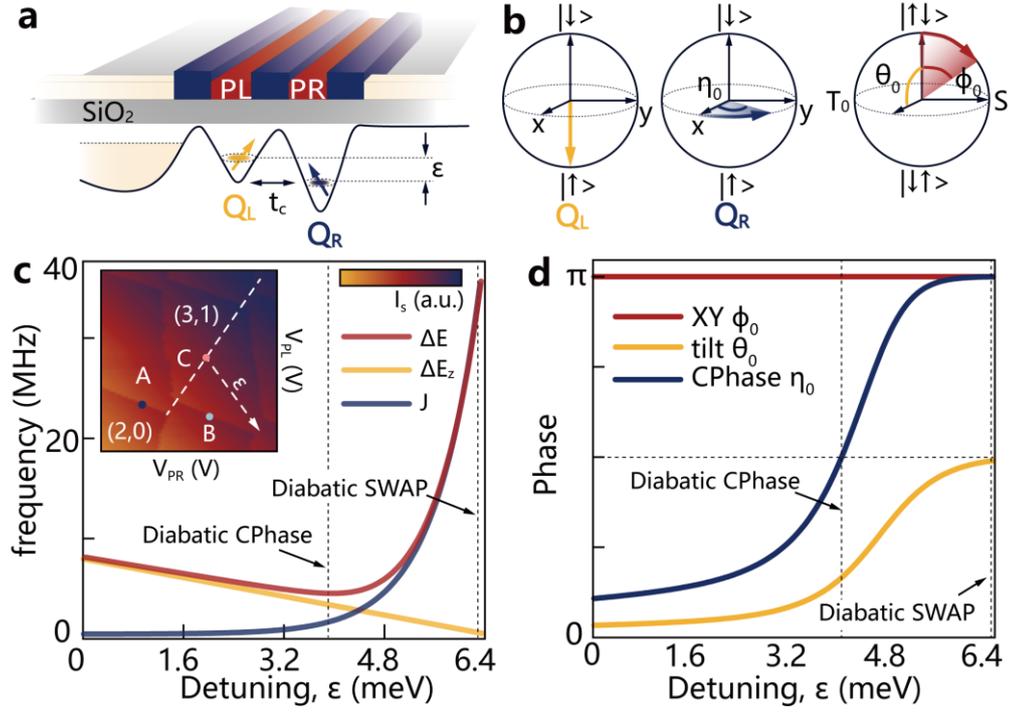

**FIG. 1. Diabatic two-qubit gates for single-spin qubits** (a) Schematic diagram of the gate structure of a Si-MOS device. The tunnel coupling $t_c$ and the detuning $\varepsilon$ between two QDs can be adjusted by tuning the electrode voltages. (b) The two-qubit operation of the diabatic pulse is divided into two components. One component is the CPhase operation, where the operation angle $\eta_0$ is represented by the blue arrow on the single-qubit Bloch sphere $Q_R$. The other component is a rotation operation around a tilt axis in the two-qubit Bloch sphere. The red arrow represents the rotation angle $\phi_0$, and the yellow solid line represents the tilt angle of the rotation axis $\theta_0$. (c) The Zeeman splitting energy difference $\Delta E_Z$, the energy difference between the antiparallel states $\Delta E$, and the exchange coupling $J$ as a function of $\varepsilon$ in our device. The insert is the charge diagram near the region (3,1). read out with SET current. Point A, B and C corresponds to the read and initial point of $Q_R$, the read and initial point of $Q_L$, and the point of symmetric operation point. The dashed arrow line points to the anticrossing between (3,1) - (2,2). (d) When the rotation angle $\phi_0 = \pi$, the simulated CPhase operation angle $\eta_0$ and the tilt angle $\theta_0$ as the function of ε. The two dashed lines in (c) and (d) indicate the operation points for the diabatic CPhase gate and the SWAP gate.

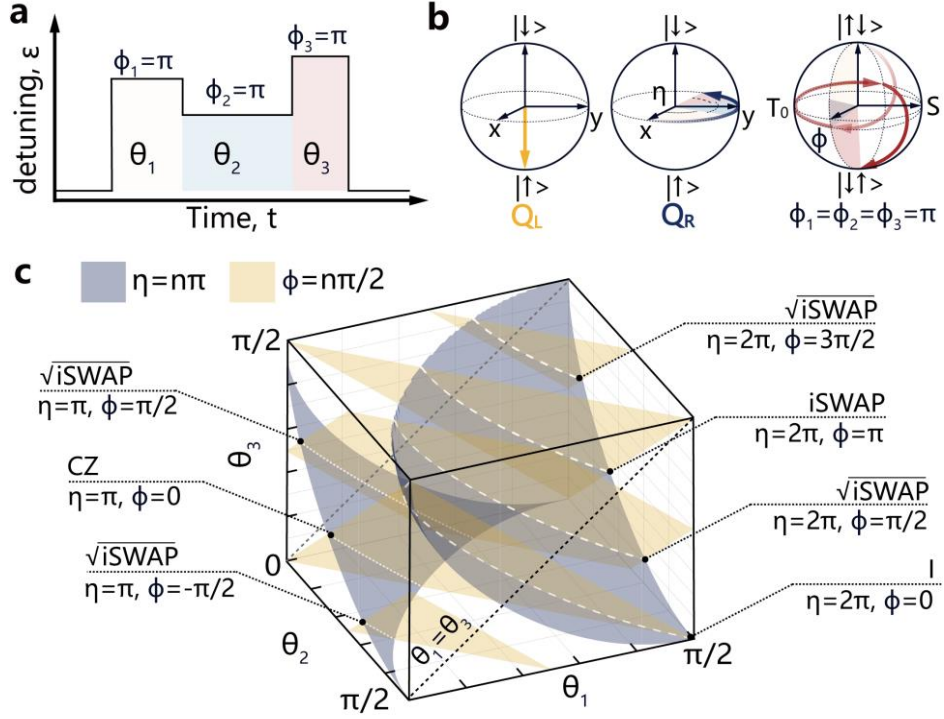

**FIG. 2. Composite gate scheme.** (a) Schematic diagram of the three-segment composite pulse. $\theta_1, \theta_2, \theta_3$ are the title angles of the rotation axis corresponding to each pulse segment. The duration of each pulse segment is tuned to ensure that $\phi_1 = \phi_2 = \phi_3 = \pi$. (b) The composition operation implemented by the three-segment diabatic pulse. $\eta$ represents the CPhase operation angle, and $\phi$ represents the rotation angle of the iSWAP operation. The fan shapes in different colors correspond to the accumulated phases in different segments. (c) The composite pulse parameters $(\theta_1, \theta_2, \theta_3)$ correspond to the specific two-qubit composite operations $(\eta, \phi)$. Each point on the purple surface $(\theta_1, \theta_2, \theta_3)$ enables the composite operation with $\eta = n\pi (n = 1,2,3 \cdots)$. Each point on the yellow surface corresponds to a composite operation with $\phi = \frac{n\pi}{2} (n = 1,2,3 \cdots)$. Each point on the intersection line of the two sets of surfaces corresponds to a three-segment pulse that applies an essential two-qubit gate operation.

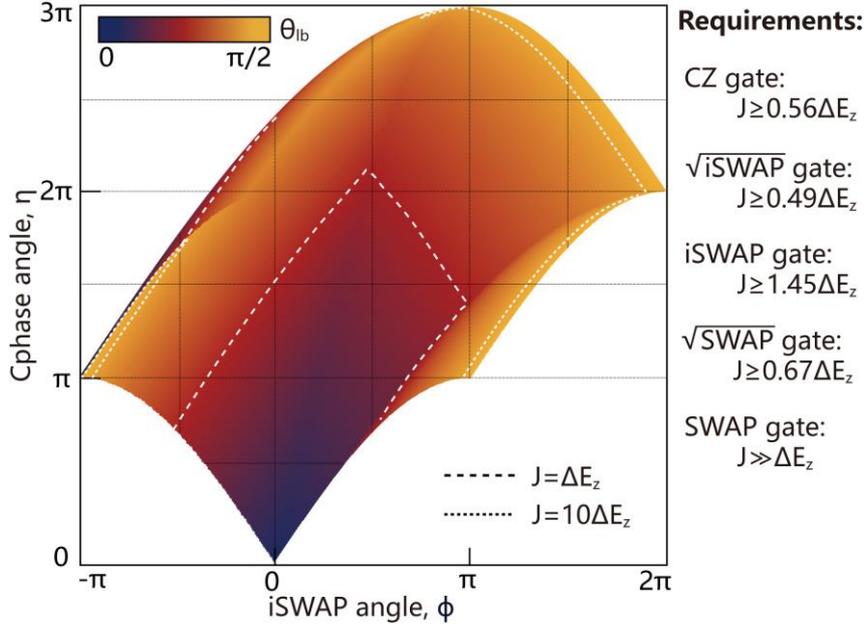

**FIG. 3. Operation angle scale.** The lower bound of the required rotation angle $\theta_{lb}$ as a function of the operation angles $(\eta, \phi)$. The colored region indicates the operation combination we can realized. The white dashed lines give the boundaries corresponding to the requirements $J = \Delta E_z$ and $J = 10\Delta E_z$. The items in the right column list the required lowest ratios $\frac{J}{\Delta E_Z}$ for realizing different essential two-qubit gates with a single composite two-qubit gate.

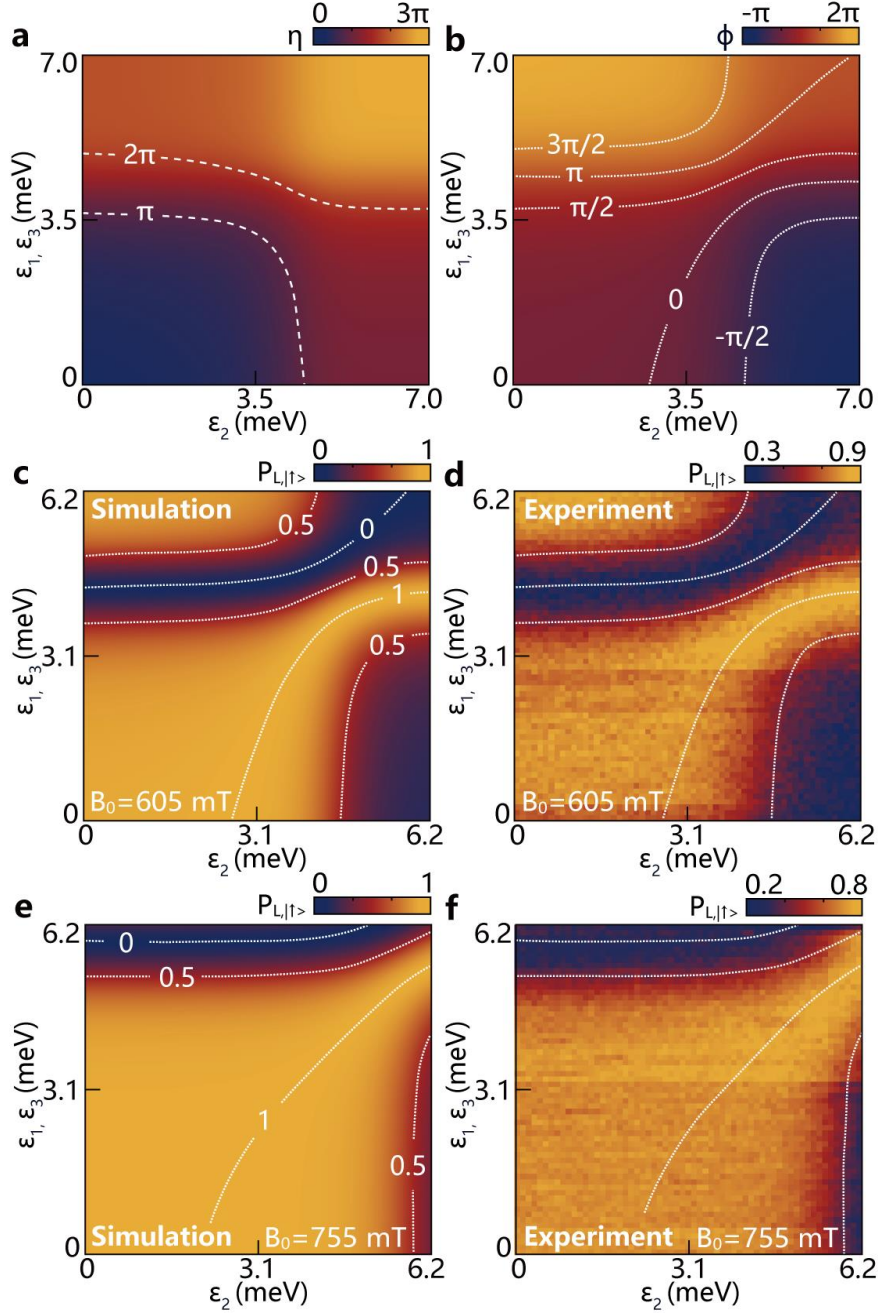

**FIG. 4. Verification of the composite gate scheme.** The dependences between $(\eta, \phi)$ and the three-segment composite pulse amplitude $(\varepsilon_1, \varepsilon_2, \varepsilon_3)$ are given in (a) and (b), respectively. We apply the composite gate to a two-qubit system with the initial state $|\uparrow\downarrow\rangle$. With $B_0 = 605$ mT, the simulated and experimentally measured spin-up probabilities of $Q_L$ $P_{L,|\uparrow\rangle}$ are given in (c) and (d), respectively. (e) and (f) correspond to the simulated and experimentally measured results at $B_0 = 755$ mT, respectively. The dashed lines in (c)-(f) illustrate the contours at $P_{L,|\uparrow\rangle} = 0$, $P_{L,|\uparrow\rangle} = 0.5$ and $P_{L,|\uparrow\rangle} = 1$.

# Supplementary Information

# A diverse set of two-qubit gates for spin qubits in semiconductor quantum dots

**Supplementary Note 1: The composite gate theory**

**Supplementary Note 2: Additional data for simulated angles at 755 mT**

**Supplementary Note 3: Gate Fidelity Simulation**

**Supplementary Note 1: The composite gate theory**

Our theoretical derivation starts from the Heisenberg Hamiltonian [1], which describes a two-qubit system

$$H_{lab} = J\left(\hat{S}_L \cdot \hat{S}_R - \frac{1}{4}\right) + \hat{S}_{zL} E_{zL} + \hat{S}_{zR} E_{zR}, \quad (S1)$$

where $J$ represents the exchange interaction between $Q_L$ and $Q_R$, and $E_{zL}$ and $E_{zR}$ describe the Zeeman energies of the two spins. The Hamiltonian in the two-qubit basis $(|\uparrow\uparrow\rangle, |\uparrow\downarrow\rangle, |\downarrow\uparrow\rangle, |\downarrow\downarrow\rangle)^T$ can be written as [1, 2]

$$H_{lab} = \begin{pmatrix} E_z & & & \\ & \frac{\Delta E_z - J}{2} & \frac{J}{2} & \\ & \frac{J}{2} & \frac{-\Delta E_z - J}{2} & \\ & & & -E_z \end{pmatrix}. \quad (S2)$$

Here, we define the average Zeeman splitting as $E_z = \frac{E_{zL} + E_{zR}}{2}$, the difference in Zeeman energy as $\Delta E_z = E_{zL} - E_{zR}$ and $h = 1$. As mentioned in the main text, a diabatic pulse on the plunger gate induces an operator

$$U = e^{-iH_{lab}t} = Z_L(\gamma_L)Z_R(\gamma_R)\begin{pmatrix} 1 & 0 & 0 & 0 \\ 0 & \cos\left(\frac{\phi_0}{2}\right) - i\cos(\theta_0)\sin\left(\frac{\phi_0}{2}\right) & -i\sin(\theta_0)\sin\left(\frac{\phi_0}{2}\right) & 0 \\ 0 & -i\sin(\theta_0)\sin\left(\frac{\phi_0}{2}\right) & \cos(\phi_0) + i\cos(\theta_0)\sin\left(\frac{\phi_0}{2}\right) & 0 \\ 0 & 0 & 0 & e^{-i\eta_0} \end{pmatrix}, \quad (S3)$$

including a CPhase operation with rotating angle $\eta_0 = Jt$ and an rotation in the antiparallel subspace, with angle $\phi_0 = \Delta E t$ around the tilt axis $\theta_0 = \arctan\left(\frac{J}{\Delta E_z}\right)$. Here, $\Delta E$ is the energy difference between the antiparallel states $|\widetilde{\uparrow\downarrow}\rangle$ and $|\widetilde{\downarrow\uparrow}\rangle$, and is calculated as $\Delta E = \sqrt{J^2 + \Delta E_z^2}$. For each segment in a composite gate, the rotation angle equals to π; thus, the duration $T = \frac{\pi}{\Delta E}$, and its evolution operator is given as

$$U_n = \begin{pmatrix} e^{-\frac{i\pi E_{z,n}}{\Delta E_n}} & & & \\ & -ie^{\frac{i\pi\sin(\theta_n)}{2}}\cos(\theta_n) & -ie^{\frac{i\pi\sin(\theta_n)}{2}}\sin(\theta_n) & \\ & -ie^{\frac{i\pi\sin(\theta_n)}{2}}\sin(\theta_n) & ie^{\frac{i\pi\sin(\theta_n)}{2}}\cos(\theta_n) & \\ & & & e^{\frac{i\pi E_{z,n}}{\Delta E_n}} \end{pmatrix}, \quad (S4)$$

where the index $n$ indicates the serial number of the pulse segment, $\cos(\theta_n) = \frac{J_n}{\sqrt{\Delta E_{z,n}^2 + J_n^2}}$ and $\sin(\theta_n) = \frac{\Delta E_{z,n}}{\sqrt{\Delta E_{z,n}^2 + J_n^2}}$.

For a composite gate corresponding to the three-segment pulse, the operator can be given as

$$U = U_3 U_2 U_1$$
$$= \begin{pmatrix} e^{-i\pi E_z\left(\frac{1}{\Delta E_1}+\frac{1}{\Delta E_2}+\frac{1}{\Delta E_3}\right)} & & & \\ & ie^{\frac{i\eta}{2}}\cos\left(\frac{\phi}{2}\right) & ie^{\frac{i\eta}{2}}\sin\left(\frac{\phi}{2}\right) & \\ & ie^{\frac{i\eta}{2}}\sin\left(\frac{\phi}{2}\right) & -ie^{\frac{i\eta}{2}}\cos\left(\frac{\phi}{2}\right) & \\ & & & e^{i\pi E_z\left(\frac{1}{\Delta E_1}+\frac{1}{\Delta E_2}+\frac{1}{\Delta E_3}\right)} \end{pmatrix} \quad (S5)$$

Here we define

$$\eta = \pi[\sin(\theta_1) + \sin(\theta_2) + \sin(\theta_3)], \quad (S6)$$

$$\phi = 2 * (\theta_1 - \theta_2 + \theta_3) \quad (S7)$$

To totally sperate single-qubit operations and the two-qubit operations, firstly, we sperate the Ising and XY operation part out. The total operator now is written as:

$$U = \begin{pmatrix} e^{-i\pi E_z \left(\frac{1}{\Delta E_1}+\frac{1}{\Delta E_2}+\frac{1}{\Delta E_3}\right)} & & & \\ & i & & \\ & & -i & \\ & & & e^{i\pi E_z \left(\frac{1}{\Delta E_1}+\frac{1}{\Delta E_2}+\frac{1}{\Delta E_3}\right)} \end{pmatrix}$$

$$\begin{pmatrix} 1 & & & \\ & e^{\frac{i\eta}{2}} & & \\ & & e^{\frac{i\eta}{2}} & \\ & & & 1 \end{pmatrix} \begin{pmatrix} 1 & & & \\ & \cos\left(\frac{\phi}{2}\right) & \sin\left(\frac{\phi}{2}\right) & \\ & -\sin\left(\frac{\phi}{2}\right) & \cos\left(\frac{\phi}{2}\right) & \\ & & & 1 \end{pmatrix}. \quad (S8)$$

For the Ising part, it can be written as the composition of single-qubit phase and CPhase operation, while the final matrix in S8 is actually an iSWAP operation with four $\frac{\pi}{4}$ single qubit phases. The derivation of the two conclusions are given as:

$$\begin{pmatrix} 1 & & & \\ & e^{\frac{i\eta}{2}} & & \\ & & e^{\frac{i\eta}{2}} & \\ & & & 1 \end{pmatrix} = e^{\frac{i\eta}{2}} \begin{pmatrix} e^{-\frac{i\eta}{2}} & & & \\ & 1 & & \\ & & 1 & \\ & & & e^{\frac{i\eta}{2}} \end{pmatrix} \begin{pmatrix} 1 & & & \\ & 1 & & \\ & & 1 & \\ & & & e^{-i\eta} \end{pmatrix} \quad (S9)$$

$$\begin{pmatrix} 1 & & & \\ & \cos\left(\frac{\phi}{2}\right) & \sin\left(\frac{\phi}{2}\right) & \\ & -\sin\left(\frac{\phi}{2}\right) & \cos\left(\frac{\phi}{2}\right) & \\ & & & 1 \end{pmatrix} = \begin{pmatrix} 1 & & & \\ & e^{\frac{i\pi}{4}} & & \\ & & e^{-\frac{i\pi}{4}} & \\ & & & 1 \end{pmatrix} \begin{pmatrix} 1 & & & \\ & \cos\left(\frac{\phi}{2}\right) & -i\sin\left(\frac{\phi}{2}\right) & \\ & -i\sin\left(\frac{\phi}{2}\right) & \cos\left(\frac{\phi}{2}\right) & \\ & & & 1 \end{pmatrix} \begin{pmatrix} 1 & & & \\ & e^{-\frac{i\pi}{4}} & & \\ & & e^{\frac{i\pi}{4}} & \\ & & & 1 \end{pmatrix} \quad (S10)$$

After well organizing single-qubit operations and the two-qubit operations, we can give the operator of the composite gate as:

$$\begin{aligned} U &= U_3 U_2 U_1 \\ &= Z_L(\gamma_L) Z_R(\gamma_R) Z_L\left(\frac{\pi}{4}\right) Z_R\left(-\frac{\pi}{4}\right) \\ &\quad \begin{pmatrix} 1 & 0 & 0 & 0 \\ 0 & \cos\left(\frac{\phi}{2}\right) & -i\sin\left(\frac{\phi}{2}\right) & 0 \\ 0 & -i\sin\left(\frac{\phi}{2}\right) & \cos\left(\frac{\phi}{2}\right) & 0 \\ 0 & 0 & 0 & e^{-i\eta} \end{pmatrix} \\ &\quad Z_L\left(-\frac{\pi}{4}\right) Z_R\left(\frac{\pi}{4}\right), \end{aligned} \quad (S11)$$

where $Z_L(\gamma_L)$ and $Z_R(\gamma_R)$ are the single-qubit local phases accumulated during the composite operation and are determined by the pulse amplitude. Here, we focus on the two-qubit operation. Then, the two-qubit operation is

$$U_{lab} = \begin{pmatrix} 1 & 0 & 0 & 0 \\ 0 & \cos\left(\frac{\phi}{2}\right) & -i\sin\left(\frac{\phi}{2}\right) & 0 \\ 0 & -i\sin\left(\frac{\phi}{2}\right) & \cos\left(\frac{\phi}{2}\right) & 0 \\ 0 & 0 & 0 & e^{-i\eta} \end{pmatrix}, \tag{S12}$$

which indicates that $\eta$ is the CPhase operation angle a and $\phi$ is the iSWAP operation angle respectively. For the five parameters of the composite operation (two operation angles and three pulse amplitude parameters), we can deduce the other two parameters by knowing only three arbitrary parameters.

With the composite gate theory, the gate operation angle $(\eta, \phi)$ in experiments can be predicted for a given composite pulse $[\theta_1(\varepsilon_1), \theta_2(\varepsilon_2), \theta_3(\varepsilon_3)]$. The simulation results are given in Fig. 4 (605 mT) and Fig. S1 (755 mT) according to Eq. S7-8. To verify these gate operations, a series of experiments are designed, and the results are also simulated.

To verify the operation angle of the iSWAP, we first initialize the two-qubit system to the $|\uparrow\downarrow\rangle$ state and then perform the composite operation before determining the two-qubit probability. The spin-up probability of $Q_L$ is $P_{L,|\uparrow\rangle} = \frac{1+\cos(\phi)}{2}$, which depends only on the operation angle of the iSWAP gate. The experimental results and the theoretical results match well, as given in the main article.

**Supplementary Note 2: Additional data for simulated angles at 755 mT**

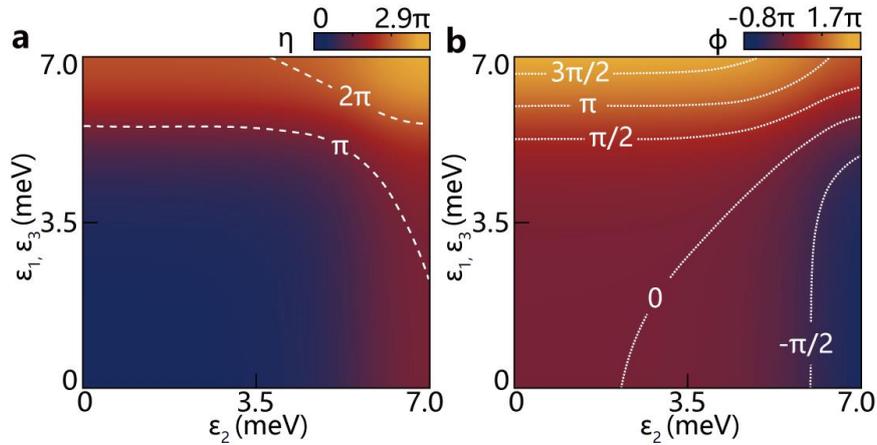

**FIG. S1.** When $B_0 = 755$ mT, the relationship between $(\eta, \phi)$ and the three-segment diabatic pulse amplitude $(\varepsilon_1, \varepsilon_2, \varepsilon_3)$ is given in (a) and (b), respectively.

## Supplementary Note 3: Gate Fidelity Simulation

With the composite pulse essentially performing a continuous set of two-qubit gates, an important question is how charge noise influences the overall gate fidelity. Here we perform a series of numerical simulations assuming that gate errors are dominated by quasi-static charge noise on $\varepsilon$ [2,3]. With crucial parameters $J, \Delta E_z$ and $\Delta E$ in our system all functions of $\varepsilon$, they are all influenced by the charge noise and extracted from the DQD spectrum. For each target composite two-qubit operation, there could be more than one available composite pulses at $B_0 = 755$ mT. During our simulation, we always chose the composite pulse which requires the minimal $J/\Delta E_z$. We first assume the charge noise is Gaussian with zero mean and generate a random noise, then calculate gate fidelity $F_{gate} = tr(U_{noise}^\dagger U_{ideal})$. Here $U_{noise}$ and $U_{ideal}$ corresponds to the operator with and without the noise effect, respectively. For each noise magnitude $\sigma_\varepsilon$, we repeat this process until the averaged fidelity is converges. Finally, we extract the noise amplitude $\sigma_\varepsilon$ threshold corresponding to the gate fidelity of 99% for each target composite two-qubit operation. The threshold of $\sigma_\varepsilon$ and the average pulse amplitude for different operation angles are given in Fig. S2(a) and (b), respectively. Due to the limited adjustable range for the ratio $J/\Delta E_z$, some of the composite operation angles are not experimentally feasible, although they could become feasible with further composite operations. The agreement between the average pulse amplitude and the noise magnitude threshold is clear. This agreement is expected because the noise sensitivity increases as the diabatic pulse amplitude increasing. In general, the noise sensitivity depends strongly on the amplitude of each pulse segment [4,5].

To further investigate the influence of $\sigma_\varepsilon$ on different two-qubit gates, we simulate various gate errors as functions of $\sigma_\varepsilon$. The corresponding results are shown in Fig. S2(c). Due to the higher average pulse amplitude and higher ratio requirement, compared to those of the CPhase, $\sqrt{iSWAP}$ and $\sqrt{SWAP}$ gates, the gate fidelities of the iSWAP gates are more susceptible to charge noise. Nevertheless, under the condition reported in our previous work $\sigma_\varepsilon \approx 16$ μeV [2], all four types of essential gates can be realized with a fidelity greater than 99%.

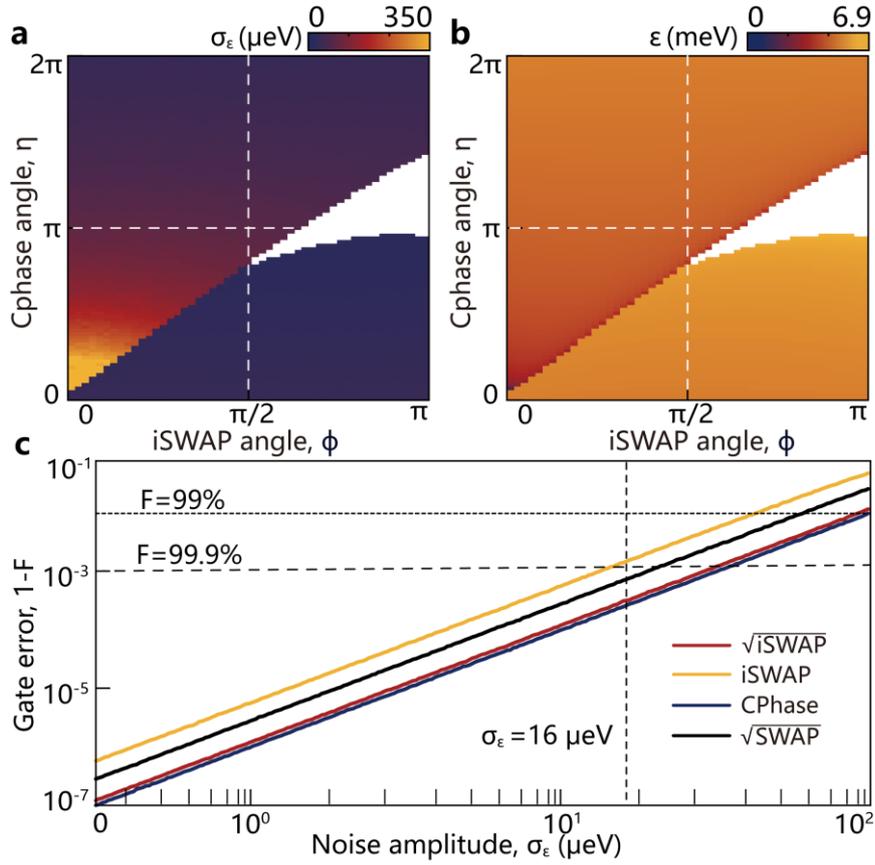

**FIG. S2. Gate fidelity and noise magnitude at $B_0 = 755$ mT.** (a) For a composite operation with gate fidelity 99%, the maximum tolerated noise amplitude for different operation angles $(\eta, \phi)$. (b) The average pulse amplitude for different composite operation angles. (c) The simulated gate errors as a function of the noise amplitude for four types of essential gates. The transverse dense dashed lines and the spaced dashed lines highlight the gate fidelity values of 99% and 99.9%, respectively. The vertical dashed line highlights the typical noise magnitude $\sigma_\varepsilon = 16$ μeV.